\documentstyle[aps,epsfig]{revtex}
\begin{document}
\preprint{\vbox{\hbox{INP MSU -- 98-24/525} \vspace{-0.3cm}
                \hbox{IFT -- P.040/98}       \vspace{-0.3cm}}}

\title{Single Top Quark at Future Hadron Colliders.\\
        Complete Signal and Background Study.}

\author{ A.\ S.\ Belyaev$^{1,2}$, E.\ E.\ Boos$^2$, L.\ V.\ Dudko$^2$}

\address{$^1$ Instituto de F\'{\i}sica Te\'orica, 
              Universidade Estadual Paulista, \\ 
              Rua Pamplona 145, 01405--900 S\~ao Paulo, Brazil.\\
         $^2$ Skobeltsyn Institute of Nuclear Physics, 
              Moscow State University \\
              119899 Moscow, Russian Federation \\} 

\maketitle

\begin{abstract}
We perform a detail theoretical study including decays and jet
fragmentation of all the important modes of
the single top quark production and all basic background
processes at the upgraded Tevatron and LHC colliders.  
Special attention was paid to the  complete tree level calculation of the QCD
fake background which was not considered   in the  previous studies. 
Analysis of the various kinematical distributions for
the signal and   backgrounds  allowed to work out a set  of
cuts for an  efficient background suppression and extraction of the
signal. It was shown that the signal to background ratio after
optimized cuts could reach about 0.4 at the Tevatron and  1 at the 
LHC. The remaining after cuts rate of the signal at the LHC  for the
$lepton+jets $ signature  is expected to be about 6.1~pb and will be enough
to study the single top physics even during the LHC operation at a low
luminosity.  
\end{abstract}

\section{Introduction}

The existence  of the top quark has been established in March 1995
by the CDF and {D\O}  collaborations at the Tevatron collider. 
\cite{cdf-d0}. Top quark has been discovered in the strong
{\mbox{${t\bar{t}}$}} pair production mode.

The cross section of  electroweak process of   single top quark
was found to be  comparable with the 
QCD pair top production\cite{st}. 
Single top production mechanism is the independent way of a confirmation
of the top quark existence and  straightforward key   
to measure the $Vtb$ CKM matrix element and to study the $Wtb$ vertex.
Since the mass of the top quark is very large compared to all other quarks
one might expect some deviations from the Standard Model (SM) predictions
in the top quark interactions \cite{peccei}.
The single top quark production rate is directly proportional to the
$Wtb$ coupling and therefore it is  a promising place to look for
deviations from the SM.

However one should stress that the task of background reduction is much
more serious and important problem in the case of the single top 
comparing to the $t\bar{t}$-pair production. 
It happens because the jet multiplicity of single top quark events 
is typically less than for $t\bar{t}$-pair production and so QCD  $Wjj$
and multijet backgrounds are much higher, and the problem of
the single top signal extraction is more involved.
That is why the detail background study is especially needed 
in order to find an optimal strategy to search for of single top quark.

Top quark decays into a $W$-boson and a $b$~quark with the 
almost 100\% branching ratio in the framework of the Standard Model. We
consider here the subsequent leptonic decays of the $W$-boson to a
electron (muon) and neutrino, as this signal has much less
background and should be easier to find
experimentally than channels with hadronic decay of the
$W$~boson.

\section{MC simulation}

In order to study a possibility of the signal extraction
from the background we have created the MC generator for complete set
of the  single top production  and backgrounds processes.
Generator  was designed as a new external user  process for
the PYTHIA~5.7/JETSET~7.4  package~\cite{pythia}.
This generator is related to  PYTHIA~5.7 by a
special interface and uses FORTRAN codes of  squared matrix elements
produced by the package CompHEP~\cite{comp}. For integration 
over the phase space and a consequent event simulation the Monte-Carlo
generator uses the kinematics with a proper smoothing of singular
variables \cite{pukhov} from the CompHEP and the integrator package
BASES/SPRING~\cite{bases}. 

The effects of the final state radiation, hadronization and string
jet fragmentation (by means of JETSET~7.4) have also been taken into
account.  The following resolutions 
 have been used for the jet and electron energy
smearing:  $\Delta E^{had}/E=0.5/\sqrt{E}$ and
$\Delta E^{ele}/E=0.2/\sqrt{E}$. 
In our analysis
we used the cone algorithm for the jet reconstruction
with the cone size 
$\Delta R= \sqrt{\Delta\varphi^2 + \Delta\eta^2}=0.5$.
 The minimum $E_T$ threshold for a
cell to be considered as a jet initiator has been chosen 5~GeV while
the one of summed $E_T$ for a collection of cells to be accepted as a
jet has been chosen 10~GeV.

For all calculations CTEQ3M parton distribution has been used.
For the top-quark production we chose
the QCD $Q^2$ scale equal to the top mass squared, while for $Wjj$
background $Q^2= {M_W}^2$ have been taken. For calculations of  
$jb\bar{b}$ and $jjb\bar{b}$ processes we chose the invariant
$b\bar{b}$ mass for the $Q^2$ scale.

Under the assumptions mentioned above the kinematic features of both
signatures for signal and background have been studied.

\subsection{Signal}

\begin{figure}[htb]
\vspace*{-2cm}
   \epsfxsize=14cm  
    \epsffile{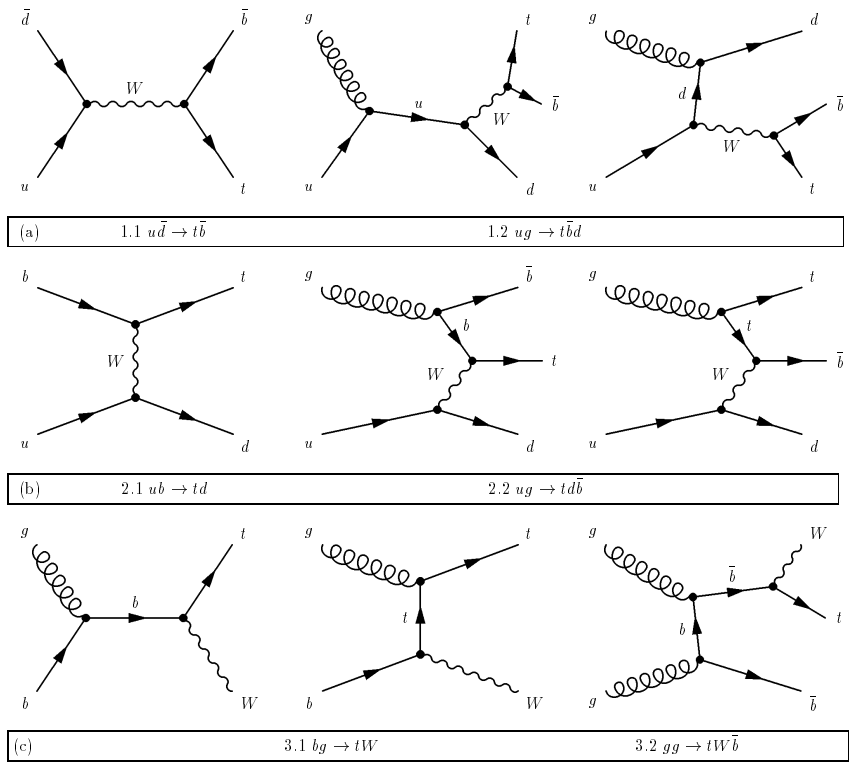}
    \vspace*{-4.0cm}
    \caption{\label{diag-st} Diagrams for single top production}   
\vspace*{-0.3cm}
\end{figure}

We concentrated on the following set of processes at Tevatron
$p\bar{p}$  and LHC $pp$ colliders leading to
the single top quark production:\\
1.~${\mbox{${p\bar{p}}$}}{\mbox{ $\rightarrow$ }}tq{\mbox{${\bar{b}}$}}+X$, \\
2.~${\mbox{${p\bar{p}}$}}{\mbox{ $\rightarrow$ }}t{\mbox{${\bar{b}}$}}+X$, \\
3.~${\mbox{${p\bar{p}}$}}{\mbox{ $\rightarrow$ }}tq+X$,\\
4.~${\mbox{${p\bar{p}}$}}{\mbox{ $\rightarrow$ }}tW +X$,\\
where $q$ is a light quark and $X$ represents  the remnants of 
the proton and antiproton. Basic Feynman diagrams for processes mentioned 
above are shown in Fig.~\ref{diag-st}. We refer to the
paper \cite{our-st} which consider the whole set of  Feynman
diagrams for signal subprocesses. 

It is necessary to stress that ${\mbox{${p\bar{p}}$}}{\mbox{ $\rightarrow$ }}tW+X$ process
contributes only with 5\% to the total cross section at the 2 TeV upgraded
Tevatron.
It could  be easily omitted at Tevatron energies but 
should be taken into account for the LHC  energies since
as we shall demonstrate below its contribution
at the LHC will be about 30\% of the  the total cross section 
of the single top quark production.

For events analysis we have rescaled the total cross sections
of the single top production
 using the results of the NLO calculations from the papers
\cite{st-nlo}($m_t$=175 GeV):\\
for $\sqrt{s}$=2 TeV\\ 
$\sigma(t\bar{b})$=$0.88\pm 0.05$ pb, $\sigma(Wg=tq\mbox{${\bar{b}}$}+tq)$= 
     $2.43 \pm 0.4$ pb;\\
and for $\sqrt{s}$=14 TeV\\
$\sigma(t\bar{b})$=$10.2\pm 0.6$ pb, $\sigma(Wg=tq\mbox{${\bar{b}}$}+tq)$=   
    $245 \pm  9$   pb.

We found  LO cross section of ${\mbox{${p\bar{p}}$}}{\mbox{ $\rightarrow$ }}tW+X$ equal to 98 pb at
the LHC 
with about 8\% uncertainty due to a choice of different structure
functions. The cross section for ${\mbox{${p\bar{p}}$}}{\mbox{ $\rightarrow$ }}tW^-+X$ was calculated
using the following way.
The process  $gg\rightarrow tW\bar b$ has been combined 
(diagram 3.2  in Fig.~\ref{diag-st}c shows the only one 
among 8 different topologies
contributed  to  this subprocesses) with 
$bg\rightarrow tW$  (diagram 3.1  in Fig.~\ref{diag-st}c ).
Similar to the $W$-gluon fusion process 
the subtraction of gluon splitting term has been made 
in order to avoid a double counting:\\
\begin{equation}
\sigma(gb + gg \rightarrow tW +X)_{real}=
\sigma(gb\rightarrow tW)+\sigma(gg\rightarrow tW\bar b)-
\sigma(g\rightarrow b\bar b \otimes gb\rightarrow tW)
\end{equation}

In the previous studies the two above subprocess $gg\rightarrow
tW\bar b$ \cite{dittmar} and $gb\rightarrow tW$ \cite{moretti}
have been considered separately which being simply added leads
to an overestimation and stronger scale dependence of the cross section.
In addition  we calculated the complete set of Feynman diagrams for 
$u\bar u + d\bar d\rightarrow tW\bar b$
subprocesses.

However one should also take into account that the
complete set of Feynman diagrams for $W t b$ processes includes $t\bar t$
pair production subprocess. The $t\bar t$ pair production
is one the background processes included separately into analysis.
Therefore its contribution has been removed from the $tW +X$ rate.
It can be done in a gauge invariant way   
by excluding the kinematical region of 
$Wb$ invariant mass around
top quark mass within 3 top decay widths as it was used
in case of the $e^+e^-$ collisions \cite{jikia-boos}:
$$   M_t+ 3\Gamma_{top} > M_{Wb}> M_t+ 3\Gamma_{top}$$.
For our calculations we used: $M_t=175$ GeV, $\Gamma_{top}=1.59$ GeV.
 
As a result, we have 91 pb for
$\sigma(gb + gg \rightarrow tW +X)_{real}$ process , while 
$\sigma(u\bar u + d\bar d\rightarrow tW\bar b + X)$ process  
gives additional 7 pb
to the total ${\mbox{${p\bar{p}}$}}{\mbox{ $\rightarrow$ }}tW^-+X$ cross section which is
therefore expected to be about 98 pb at LHC. 

Thus the total single top rate is expected to be about
3.5 pb at the upgraded Tevatron and 350 pb at the LHC to be
compared to the $t \bar t$ production rate\cite{tt} 7.6 pb and 760 pb
respectively. Relative
contributions of different subprocesses to 
the total single top quark production cross section
at Tevatron
are the following :\\
${\mbox{${p\bar{p}}$}}{\mbox{ $\rightarrow$ }}tq\mbox{${\bar{b}}$}$(39\%),\\
${\mbox{${p\bar{p}}$}}{\mbox{ $\rightarrow$ }}tb$(30\%),\\
${\mbox{${p\bar{p}}$}}{\mbox{ $\rightarrow$ }}tq$(26\%),\\
${\mbox{${p\bar{p}}$}}{\mbox{ $\rightarrow$ }}tW+X$ (5\%).\\
Relative contribution of  ${\mbox{${p\bar{p}}$}}{\mbox{ $\rightarrow$ }}tb$ and ${\mbox{${p\bar{p}}$}}{\mbox{ $\rightarrow$ }}tW+X$ 
turn over at LHC:\\
${\mbox{${p\bar{p}}$}}{\mbox{ $\rightarrow$ }}tq\mbox{${\bar{b}}$}$(44.2\%),\\
${\mbox{${p\bar{p}}$}}{\mbox{ $\rightarrow$ }}tb$(2.8\%),\\
${\mbox{${p\bar{p}}$}}{\mbox{ $\rightarrow$ }}tq$(25\%),\\
${\mbox{${p\bar{p}}$}}{\mbox{ $\rightarrow$ }}tW+X$ (28\%).

As it was mentioned we consider here the leptonic decay modes of $W$-boson
from the top quark and therefore
the final state signature to search for the signal will be:\\ 
$e^\pm(\mu^\pm)+ \mbox{$\rlap{\kern0.25em/}E_T$} +2(3)jets$ \\
 where  one of  the jets is the  $b$-quark jet from the top-quark decay.

\subsection{Main backgrounds}

The main backgrounds leading to the same $e^\pm(\mu^\pm)+ \mbox{$\rlap{\kern0.25em/}E_T$} +2(3)jets$  
final state signature as the single top signal
are the following: 
$p\bar{p}{\mbox{ $\rightarrow$ }}W+2(3)jets$ 
(gluonic, $\alpha\alpha_s$  order and electroweak $\alpha^2$ order)),
$p\bar{p}{\mbox{ $\rightarrow$ }}t\bar{t}$ pair top quark production  and  
$j(j)b\bar{b}$ QCD fake background when  one jet
imitates the electron.

All numbers  for $Wjj(Wb\bar{b})$ and $j(j)b\bar{b}$ backgrounds are presented
below  for  the initial general cuts on jets: \\
$\Delta R_{jj(ej)} >0.5$, $p_t jet > $ 10 GeV for Tevatron;\\
$\Delta R_{jj(ej)} >0.5$, $p_t jet > $ 20 GeV for LHC.

Total cross section of  W+2jets 'gluonic' background is more than 2 
orders of magnitude higher  than the signal one.  This process
includes 32 subprocesses for $u,d$-quarks and gluons in the initial
state \cite{wjj}, and  the total cross section equals to 1240~pb for 
Tevatron and 7500~pb for LHC ($s$- and $c$-
 sea quarks give additional 3\% contribution to the total cross section).

  \begin{figure}[htb]
  \vspace*{-3.0cm}
  \hspace*{-0.5cm}
  \epsfxsize=16cm  
  \epsffile{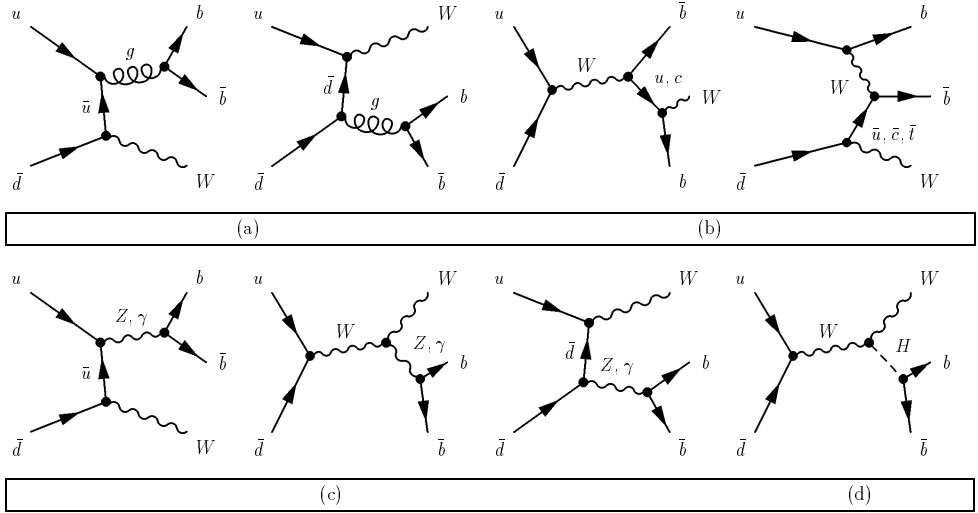}
  \vspace*{-9.0cm}
  \caption{\label{back1} Diagrams for $Wb\bar{b}$ background}
  \end{figure}

The specific feature of the single top production is the high energetic
$b$-jet in final state and  one additional $b$-jet for $W$-gluon and $W^*$
processes. It is clear that the only chance to extract the signal
from such an overwhelming background is the efficient b-quark
identification. We assume 50\% of a double b-tagging efficiency hereafter.

However the cross section of the $W+2jet$ process is so large that 
even with the requirement of a double b-tagging but due to a b-quark jet
misidentification  it represents an important part of the total 
background. In our study  we chose 0.5\% misidentification 
probability which based on  the previous MC analysis~\cite{tev33}.

The $b$-quark content of the $W+2jet$ processes is fairly small --
less then 1\%. For  the cuts mentioned above  the total cross
section  for $W^\pm b\bar{b}$  process(gluon splitting) is 8.7 pb
for Tevatron and 30 pb for LHC. However, the $W+2 b-jet$ process
is the irreducible part of the total background which 
has different kinematical properties from the main QCD $W+2jet$ part, and 
as will be shown below it depends differently on the cuts. 

Therefore the $W^\pm b\bar{b}$ background has been considered 
separately and we have calculated it completely.

Complete set of Feynman background diagrams for $Wb\bar{b}$ final state
is shown in 
Fig.~\ref{back1} for $u-$ and $\bar{d}$-quarks in the initial state.
The main contribution comes from subprocess (a) with the gluon splitting
into the $b\bar b$ quark pair.
Diagrams with virtual photon (c)  
contribute only 1\%  to the total cross section.
Contribution from $WZ$ process  (c) can be suppressed
by applying cut on the invariant $b\bar{b}$-mass.
LO cross section for $WZ$ is 2.5~pb  at Tevatron and about 30 pb at LHC. 
For our analysis we apply K-factor=1.33 (1.55) for Tevatron(LHC)
to rescale this number for the NLO total cross section\cite{wz}.
Cross section of the  process  (d)
of  Higgs production (we take as an example $m_H = $ 110 GeV) is
even smaller: 0.16 (1.8) pb (LO) at Tevatron (LHC), and it means the Higgs
is really not an important background for the single top.
We apply K-factor=1.25 (1.1) for Tevatron(LHC) to rescale the 
results for NLO one\cite{wh}. 
Diagrams  (b) give very  small contribution due to
small value of CKM elements. 

\begin{table}[htb]
\begin{center}
\begin{tabular}{ | l |  l  | l  | }
\hline
 process  	& Tevatron (pb) 		   	& LHC (pb) \\
\hline
$gg\to gb\bar{b}$			
		&  $1.64\cdot 10^5$ 			&$3.91\cdot 10^5$\\
$g\bar{u}(\bar{u}g)\to \bar{u}b \bar{b}$	
		&  $1.80\cdot 10^4(2.50\cdot 10^3)$ 	&$5.61\cdot 10^3(5.61\cdot 10^3)$\\
$g    {u}(    {u}g)\to     {u}b \bar{b}$	
		&  $2.50\cdot 10^4(1.80\cdot 10^4)$ 	&$2.41\cdot 10^4(2.41\cdot 10^4)$\\
$g\bar{d}(\bar{d}g)\to \bar{d}b \bar{b}$	
		&  $9.00\cdot 10^3(3.21\cdot 10^3)$ 	&$6.61\cdot 10^3(6.61\cdot 10^3)$\\
$g\bar{s}(\bar{s}g)\to \bar{s}b \bar{b}$	
		&  $1.97\cdot 10^3(1.97\cdot 10^3)$ 	&$4.49\cdot 10^3(4.49\cdot 10^3)$\\
$g    {d}(    {d}g)\to     {d}b \bar{b}$	
		&  $3.21\cdot 10^3(9.00\cdot 10^3)$ 	&$1.38\cdot 10^4(1.38\cdot 10^4)$\\
$g    {s}(    {s}g)\to     {s}b \bar{b}$	
		&  $1.97\cdot 10^3(1.97\cdot 10^3)$ 	&$4.49\cdot 10^3(4.49\cdot 10^3)$\\
$d\bar{d}(\bar{d}d)\to g      b \bar{b}$	
		&  $5.67\cdot 10^2(1.25\cdot 10^2)$ 	&$2..82\cdot 10^2(2.82\cdot 10^2)$\\
$u\bar{u}(\bar{u}u)\to g      b \bar{b}$
		&  $1.31\cdot 10^3(8.61\cdot 10^1)$ 	&$4.16\cdot 10^2(4.16\cdot 10^2)$\\
\hline
Total &   $2.40\cdot 10^5$ pb    &  $5.11\cdot 10^5$ pb\\
\hline  
\end{tabular}
\end{center}
\caption{Total cross section for 
$jb\bar b $ process for Tevatron and LHC. The following cuts  have been applied
at the parton level calculations:
$\Delta R_{jj} >0.5$, $p_t jet > $ 10 GeV for Tevatron and $\Delta R_{jj(ej)} >0.5$, 
$p_t jet > $ 20 GeV for LHC\label{tab:jbb}}

\end{table}

An important background is top-quark pair production:
when one of the top decays hadronically and another one --  leptonically.
One of the cut which helps to reduce the background is the cut on the
number of jets which was required to be  less than four. 
At the parton level this cut reduces
the top pair rate very strongly. However at  the simulation level with
a hadronization and jet reconstruction being taken into account 
the reduction of this background is
not so strong anymore. And as a result the  top-quark pair represents
an important part of the background. This fact will be shown below. 
NLO total cross section for 
tt-pair production was taken \cite{tt} at Tevatron to be equal to
$7.56$ pb and
760 pb at LHC.

\begin{table}[tb]
\begin{center}
\begin{tabular}{|  l | l | l | }
\hline
 process  	& Tevatron (pb) 		   	& LHC (pb) \\
\hline
$u	u	\to u	u	b \bar{b}$			
		&  $1.23\cdot 10^2$ 			 &$1.17\cdot 10^3$\\
$u\bar{u}(\bar{u}u)	\to b\bar{b}	b \bar{b}$	
		&  $2.55\cdot 10^0$ {-----------}	 &$1.06\cdot 10^0$($1.06\cdot 10^0$)\\
$u\bar{u}(\bar{u}u)	\to s\bar{s}	b \bar{b}$	
		&  $6.61\cdot 10^0$ {-----------}	 &$2.53\cdot 10^0$($2.53\cdot 10^0$)\\
$u\bar{u}(\bar{u}u)	\to c\bar{c}	b \bar{b}$	
		&  $6.52\cdot 10^0$ {-----------}	 &$2.53\cdot 10^0$($2.53\cdot 10^0$)\\
$u\bar{u}(\bar{u}u)	\to d\bar{d}	b \bar{b}$	
		&  $6.66\cdot 10^0$ {-----------}	 &$2.52\cdot 10^0$($2.53\cdot 10^0$)\\
$u\bar{u}(\bar{u}u)	\to u\bar{u}	b \bar{b}$	
		&  $8.70\cdot 10^2$ {-----------}	 &$3.38\cdot 10^2$($3.38\cdot 10^2$)\\
$u\bar{u}(\bar{u}u)	\to g   g	b \bar{b}$	
		&  $2.15\cdot 10^2$ {-----------}	 &$8.92\cdot 10^1$($8.92\cdot 10^1$)\\
$u d (du)	        \to u   d	b \bar{b}$			
		&  $1.44\cdot 10^2$($4.20\cdot 10^1$) 	 &$7.40\cdot 10^2$($7.40\cdot 10^2$)\\
$u s (su)	        \to u   s	b \bar{b}$			
		&  $9.63\cdot 10^1${-----------}	 &$1.71\cdot 10^2$($1.71\cdot 10^2$)\\
$u\bar{d}(\bar{d}u)	\to u\bar{d}   	b \bar{b}$			
		&  $3.73\cdot 10^2$($1.20\cdot 10^2$) 	 &$3.74\cdot 10^2$($3.74\cdot 10^2$)\\
$u \bar s (\bar s u)	\to u   \bar s	b \bar{b}$			
		&  $9.63\cdot 10^1${-----------}	 &$1.71\cdot 10^2$($1.71\cdot 10^2$)\\
$d\bar{u}(\bar{u}d)	\to d\bar{u}	b \bar{b}$			
		&  $3.73\cdot 10^2$($1.20\cdot 10^2$) 	 &$1.78\cdot 10^2$($1.78\cdot 10^2$)\\
$s \bar u(\bar u s)\to \bar u   s	b \bar{b}$			
		&  $9.63\cdot 10^1${-----------}  	& {-----------}  {-----------}                    \\
$\bar{u}\bar{u}	\to \bar{u}\bar{u}   	b \bar{b}$			
		&  $9.10\cdot 10^1$ {-----------}       & {-----------}  {-----------}                   \\
$\bar{u}\bar{d}(\bar{d}\bar{u})	\to \bar{u}\bar{d}   	b \bar{b}$			
		&$4.20\cdot 10^1$($1.44\cdot 10^2$)      &{-----------} {-----------}\\
$\bar u \bar s (\bar s \bar u)\to \bar u   \bar s	b \bar{b}$			
		&{-----------} ($9.63\cdot 10^1$)	& {-----------}	{-----------}		    \\
$d	d	\to d	d	b \bar{b}$			
		&  $6.40\cdot 10^1$ 			&$1.17\cdot 10^3$ \\
$d\bar{d}(\bar{d}d)	\to b\bar{b}   	b \bar{b}$			
		&  $9.38\cdot 10^{-1}$($9.38\cdot 10^{-1}$) 	&$7.00\cdot 10^{-1}$($7.00\cdot 10^{-1}$)\\
$d\bar{d}(\bar{d}d)	\to s\bar{s}   	b \bar{b}$			
		&  $2.40\cdot 10^0$($2.40\cdot 10^0$) 	&$1.70\cdot 10^{0}$($1.70\cdot 10^{0}$)\\
$d\bar{d}(\bar{d}d)	\to c\bar{c}   	b \bar{b}$			
		&  $2.35\cdot 10^0$($2.35\cdot 10^0$) 	&$1.70\cdot 10^{0}$($1.70\cdot 10^{0}$)\\
$d\bar{d}(\bar{d}d)	\to d\bar{d}   	b \bar{b}$			
		&  $2.24\cdot 10^2$($2.24\cdot 10^2$) 	&$2.05\cdot 10^{2}$($2.05\cdot 10^{2}$)\\
$d\bar{d}(\bar{d}d)	\to u\bar{u}   	b \bar{b}$			
		&  $2.40\cdot 10^0$($2.40\cdot 10^0$) 	&$1.70\cdot 10^{0}$($1.70\cdot 10^{0}$)\\
$d\bar{d}(\bar{d}d)	\to g g   	b \bar{b}$			
		&  $7.30\cdot 10^1$($7.30\cdot 10^1$) 	&$5.86\cdot 10^{1}$($5.86\cdot 10^{1}$)\\
$\bar{d}\bar{d}	\to \bar{d}\bar{d}   	b \bar{b}$			
		&  $5.10\cdot 10^1$             	& {-----------}                   \\
$d\bar{s}(\bar{s}d)	\to d\bar{s}   	b \bar{b}$			
		&  $4.23\cdot 10^1$ {-----------}	&$9.16\cdot 10^{1}$($9.16\cdot 10^{1}$)\\
$d {s}( {s}d)	\to d {s}   	b \bar{b}$			
		&  $4.23\cdot 10^1$ {-----------}	&$9.16\cdot 10^{1}$($9.16\cdot 10^{1}$)\\
$\bar{d} {s}( {s}\bar{d})	\to \bar{d}{s}   	b \bar{b}$			
		&  {-----------} $4.23\cdot 10^1$	&$9.16\cdot 10^{1}$($9.16\cdot 10^{1}$)\\
$\bar{d} {\bar{s}}( \bar{s}\bar{d})	\to \bar{d}\bar{s}   	b \bar{b}$			
		&  {-----------} $4.23\cdot 10^1$	&$9.16\cdot 10^{1}$($9.16\cdot 10^{1}$)\\
$g{u}({u}g)	\to g u   	b \bar{b}$			
		&  $7.28\cdot 10^2$($7.82\cdot 10^3$) 	&$2.37\cdot 10^{4}$($2.37\cdot 10^{4}$)\\
$g\bar{u}(\bar{u}g)	\to g \bar{u}   	b \bar{b}$			
		& $7.82\cdot 10^3$($7.28\cdot 10^2$) 	&$4.53\cdot 10^{3}$($4.53\cdot 10^{3}$)\\
$g{d}({d}g)	\to g d   	b \bar{b}$			
		&  $1.01\cdot 10^3$($3.52\cdot 10^3$) 	&$1.29\cdot 10^{4}$($1.29\cdot 10^{4}$)\\
$g{s}({s}g)	\to g s   	b \bar{b}$			
		&  $7.61\cdot 10^2$($7.61\cdot 10^2$) 	&$2.43\cdot 10^{3}$($2.43\cdot 10^{3}$)\\
$g\bar{d}(\bar{d}g)	\to g \bar{d}   	b \bar{b}$			
		& $3.52\cdot 10^3$($1.01\cdot 10^3$) 	&$5.47\cdot 10^{3}$($5.47\cdot 10^{3}$)\\
$g{\bar s}({\bar s}g)	\to g \bar s   	b \bar{b}$			
		&  $7.61\cdot 10^2$($7.61\cdot 10^2$) 	&$2.43\cdot 10^{3}$($2.43\cdot 10^{3}$)\\
$g	g	\to b	\bar{b}	b \bar{b}$			
		&  $9.90\cdot 10^1$ 			&$6.58\cdot 10^2$ \\
$g	g	\to s	\bar{s}	b \bar{b}$			
		&  $4.80\cdot 10^2$ 			&$2.21\cdot 10^3$ \\
$g	g	\to c	\bar{c}	b\bar{b}$			
		&  $4.60\cdot 10^2$ 			&$2.24\cdot 10^3$ \\
$g	g	\to d	\bar{d}	b \bar{b}$			
		&  $3.85\cdot 10^2$ 			&$2.11\cdot 10^3$ \\
$g	g	\to u	\bar{u}	b \bar{b}$			
		&  $3.85\cdot 10^2$ 			&$2.11\cdot 10^3$ \\
$g	g	\to g	\bar{g}	b \bar{b}$			
		&  $3.62\cdot 10^4$ 			&$2.43\cdot 10^5$ \\
\hline
Total &   $7.01\cdot 10^4$ pb    &  $3.62\cdot 10^5$ pb\\  
\hline
\end{tabular}
\end{center}
\caption{ Total cross section for 
$jjb\bar{b}$ process for Tevatron and LHC.
 The following cuts  have been applied
at the parton level calculations:
$\Delta R_{jj} >0.5$, $p_t jet > $ 10 GeV for Tevatron and $\Delta R_{jj(ej)} >0.5$, 
$p_t jet > $ 20 GeV for LHC
\label{tab:jjbb}}
\end{table}

Another kind of  the important reducible background comes from 
the multijet QCD processes. This happens due to a possible
misidentification of jet as an electron in the detector. Though the
probability of that is
 very small (of order of 0.01-0.03\%)\cite{fake-d0}, the cross
section for such processes is huge and  give a significant contribution
to the background for single top production.
For the analysis we took $\epsilon_{fake}$=0.02\%.
 
  We have calculated
the total cross section and made MC simulation  for $jbb$ and $jjbb$
processes  which are relevant for our signature if one the light jet 
imitates electron. 
In such a way we could simulate the basic distributions of the expected
fake background and understand how strong it could be suppressed by 
kinematical cuts. The MC simulation of the fake
background for the single top study is presented for a first time.

The background from the light jets is 
appeared to be less important despite on the fact
the light jet cross sections even with cuts are very large.
The light jet background is suppressed by three small factors,
by double mistag probability to identify light jet as a b-jet
and by the small fake probability to identify light jet as an electron.  
For instance the $gg\rightarrow ggg$ subprocess
could contribute to the background  when two gluon jets fake b-quark jets
and a third gluon jet fakes the lepton.
The cross section of  $gg\rightarrow ggg$ itself is huge but 
as was mentioned it's
contribution to the single top background is suppressed by
fake probability of gluon multiplied by double mistag probability
of two gluons. For example, the cross section of 
triple gluon production Tevatron 
is $2.7\cdot 10^7$ pb, double mistag gluon probability
is equal to $10^{-6}$, fake probability of gluon is of order $10^{-4}$.
Therefore the contribution from $gg\rightarrow ggg$ process 
is estimated to be equal to
$\simeq 10^{-2} \ pb$ 
( we use combinatoric factor 3 here) 
and one can neglect it. 

In the Table~\ref{tab:jbb} and Table~\ref{tab:jjbb} 
all subprocesses giving $jbb$ and $jjbb$  final state signature 
are shown respectively
with the corresponding cross sections for the Tevatron and LHC.
In our calculations we neglected  the double sea quark and
$c$-sea quark small contributions.
 Total cross section for $jbb$ at the Tevatron (LHC)
is 240 and 70 (511  and 362)~nb for $jbb$ and $jjbb$.

The cross section of the jbb process in the Table~\ref{tab:jbb} 
is only about 2 times higher at LHC than at Tevatron 
because higher  jet $p_T$
cuts for  LHC have been used (20 GeV at LHC and 10 GeV
at Tevatron).
If the  equal jet $p_T$ cuts are used the cross section at LHC  
is about 50 times higher  than that at Tevatron. 

We performed two ways of calculation  of the $jjb\bar b$ process,
the complete tree level calculation and the 
splitting approximation when one uses the complete result from $jb\bar b$
with an additional jet radiation from the initial and final states. 
In such a way we have checked the validity of  splitting 
approximation.

As it was expected the splitting approximation works reasonably well
for the total rate if rather soft cuts on the additional jet are used
and the difference increases if the more strong cuts are applied.
The Table~\ref{tab:split} illustrates such a difference 
in results for the approximation and exact calculations 
for various cuts on the $p_T$ of the second jet
(that is the light jet with the smallest $p_T$ which 
is more likely an additional radiated light jet).
\begin{table}
\begin{center}
\begin{tabular}{| l | l | l | l | l | }
\hline
\hline
$p_{j2T} [GeV]$& 10 & 15 & 20 & 40\\
\hline
$\sigma_{jjbb}^{exact}[pb]$
              & 70& 32& 14& 1.2\\ 
$\sigma_{jjbb}^{split}[pb]$
              & 64& 22& 8&  0.25\\
\hline
\end{tabular}
\end{center}             
\caption{Comparison the cross sections for  $jjb\bar b$ process
         for exact calculations and the splitting approximation
         for various $p_{j2T}$ cuts at Tevatron. 
         The following cuts  have been applied
at the MC level:
$\Delta R_{jj} >0.5$, $p_t$ of the first jet $> $ 10 GeV \label{tab:split}}
\end{table}
Indeed one can see that for  $p_{j2T} > 10$~GeV cut  the difference
between
exact calculation and the splitting approximation is only 
about 10 \%: 70 nb and 64 nb respectively. 
But after  $p_{j2T} > 40$~GeV  cut those  results differ
almost by factor 5: one has  1.2 and 0.25 nb for exact calculation of $jjb\bar b$
and  splitting  approximation respectively.

\begin{figure}[htb]
   \vspace*{-5.5cm}
   \begin{center}
   \begin{picture}(600,500)(0,0)
   \vspace*{-0.6cm}
   \hspace*{-0.5cm}
   \epsfxsize=12cm\epsffile{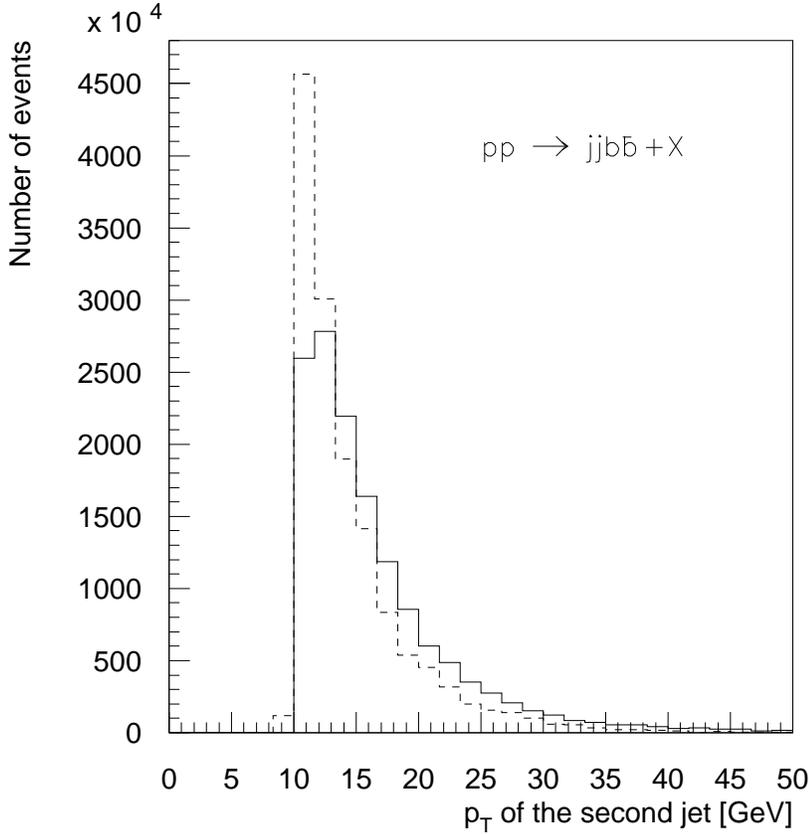}
   \end{picture}
   \end{center}
   \vspace*{-1.5cm}
   \caption{\label{split} Distribution for $p_T$ of the second jet
   for $jjb\bar b$ process for exact calculation (solid line)
    and for splitting from $jb\bar b$ process (dashed line) at
    Tevatron.} 
\end{figure}

The expected difference in the distribution on the momenta
transverse of the second jet is illustrated in  Figure~\ref{split}.
The distribution in case of the splitting approximation
is significantly softer. 

Since we do not apply high $p_T$ cut on jet (one of them fakes electron,
for which we apply 15 GeV $p_T$ cut) the difference between exact
calculation and the approximation is of order of 25\% for the
fake background simulation.

\subsection{Signal and background kinematical properties}

The rate of the signal and backgrounds  presented above clearly
shows that even
after  $b$-tagging the signal is still more than one order
less than the background.
This fact requires a special kinematical analysis 
in order to find out a strategy how to suppress the background
and extract the signal in an optimal way.

\begin{figure}[htb]
   \begin{center}
    \vskip -0.8cm\hspace*{-0.5cm}
    \epsfxsize=8cm\epsffile{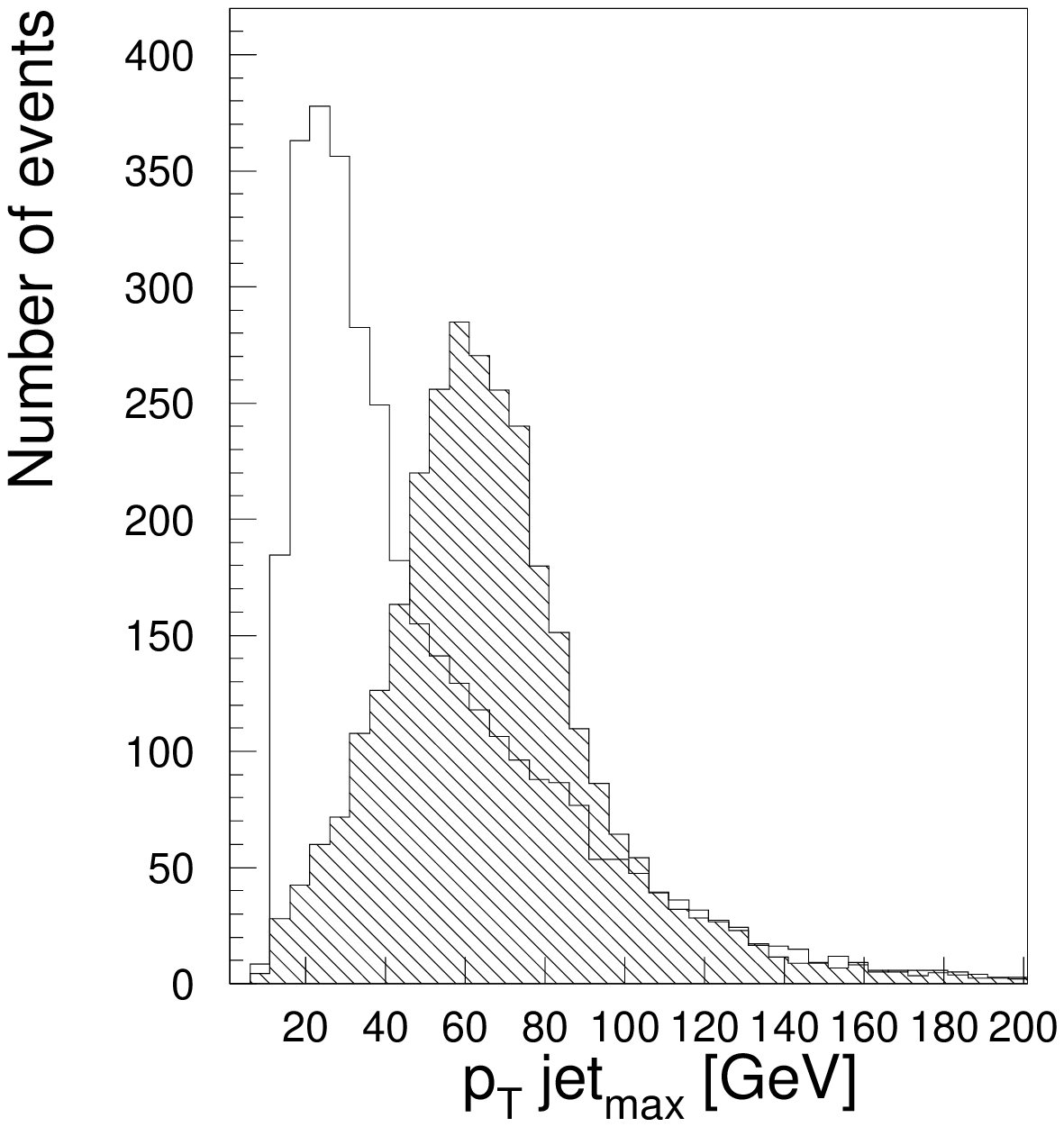}\epsfxsize=8cm\epsffile{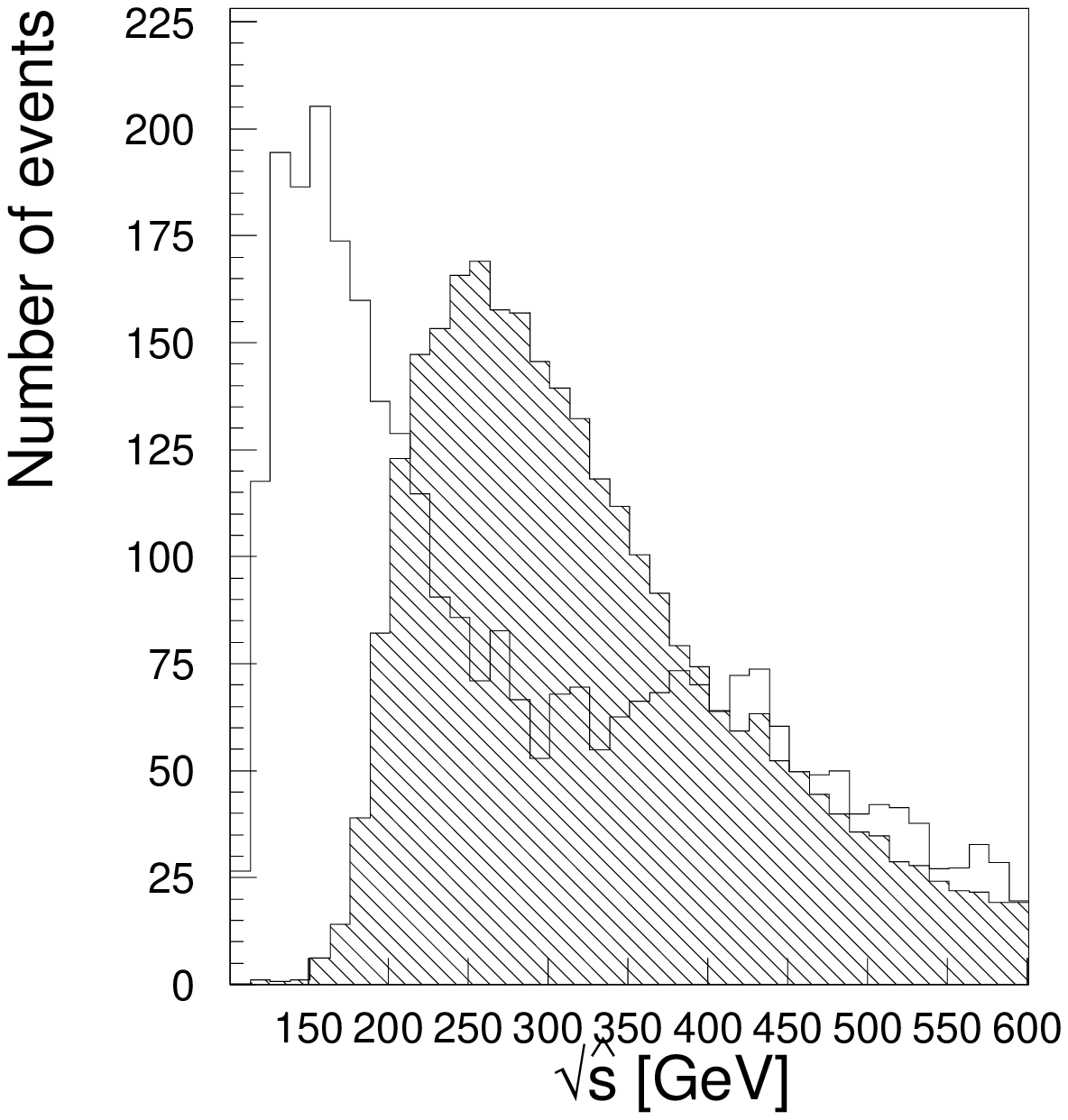}
    \vskip -1.2cm\hspace*{-0.5cm}
    \epsfxsize=8cm\epsffile{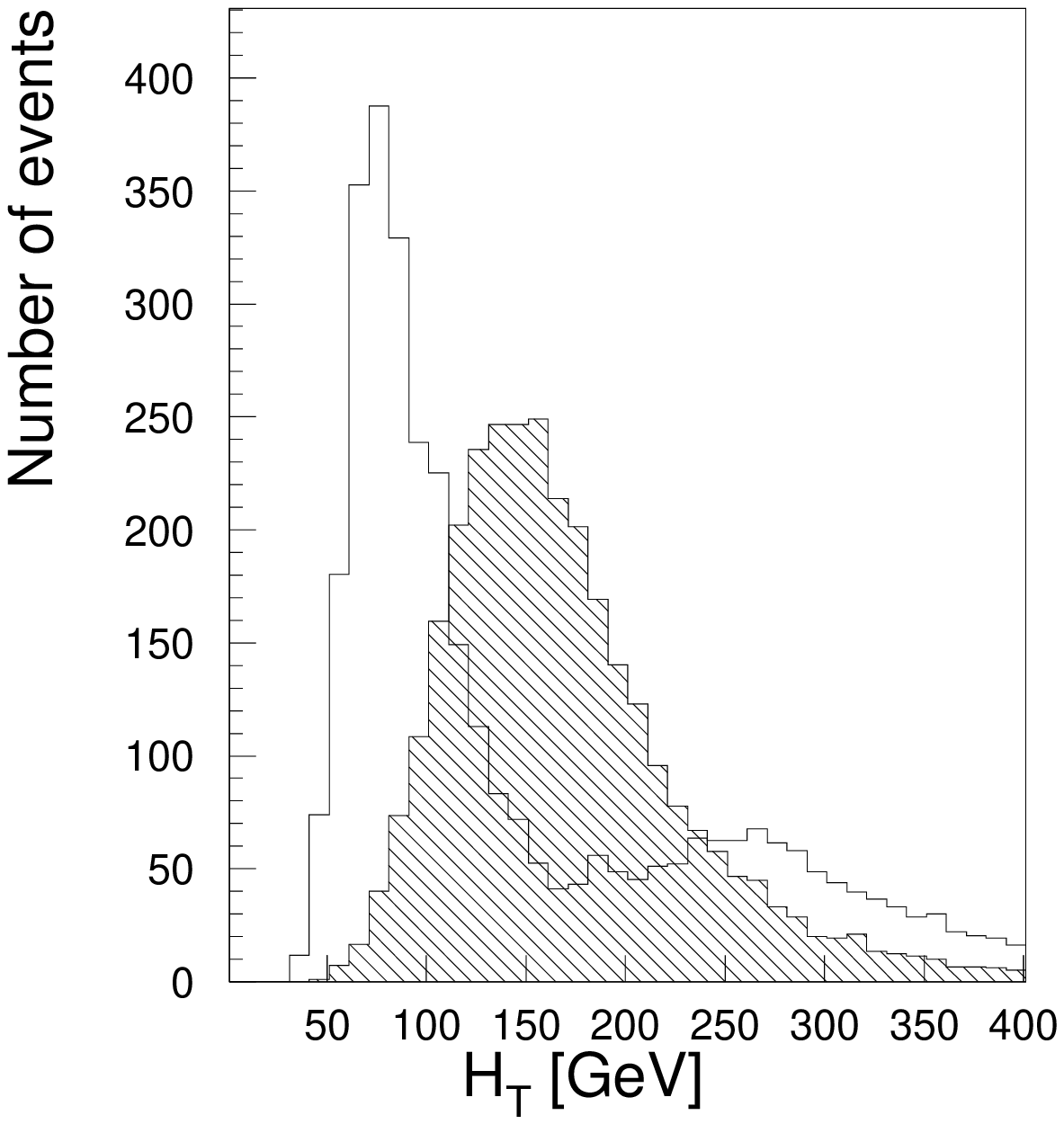}\epsfxsize=8cm\epsffile{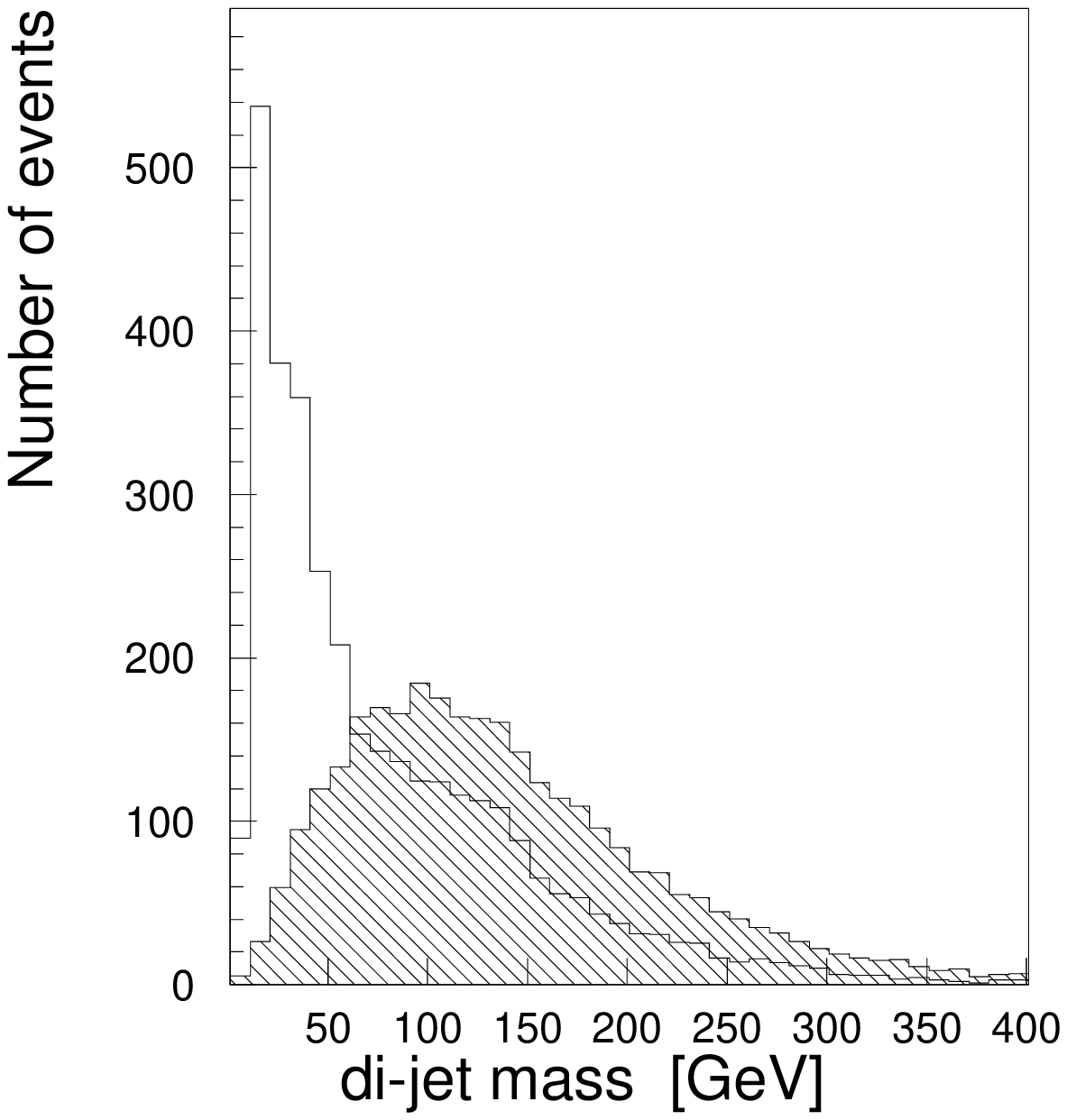}\\
  \end{center}
  \vspace*{-1.0cm}
  \caption{\label{st-fig} Distributions for signal and background
  for the some most spectacular  variables at Tevatron. Sketched histogram
  stands for signal.} 
\end{figure}

 \begin{figure}[htb]
   \begin{center}
    \vskip -0.8cm\hspace*{-0.5cm}
    \epsfxsize=8cm\epsffile{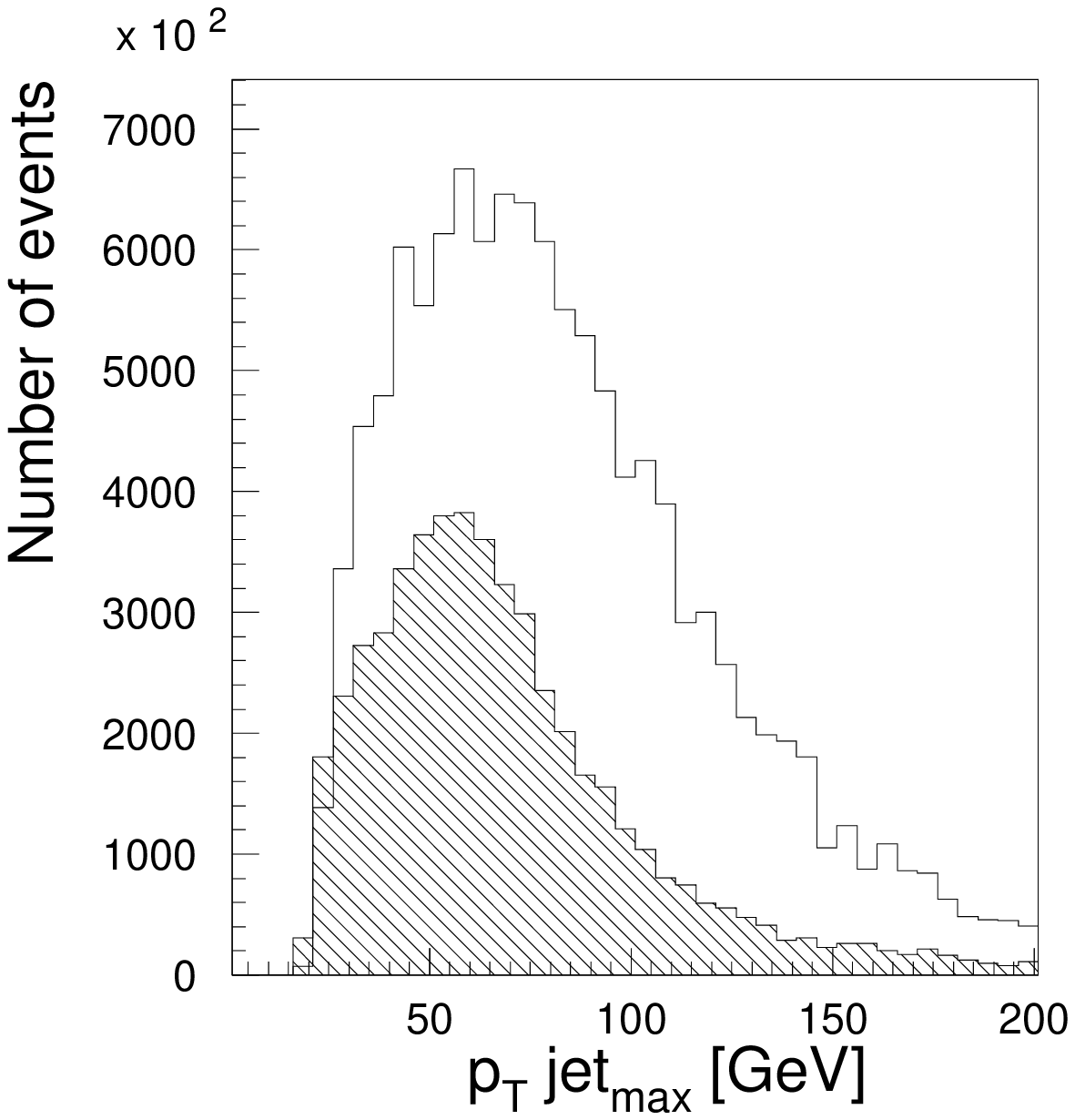}\epsfxsize=8cm\epsffile{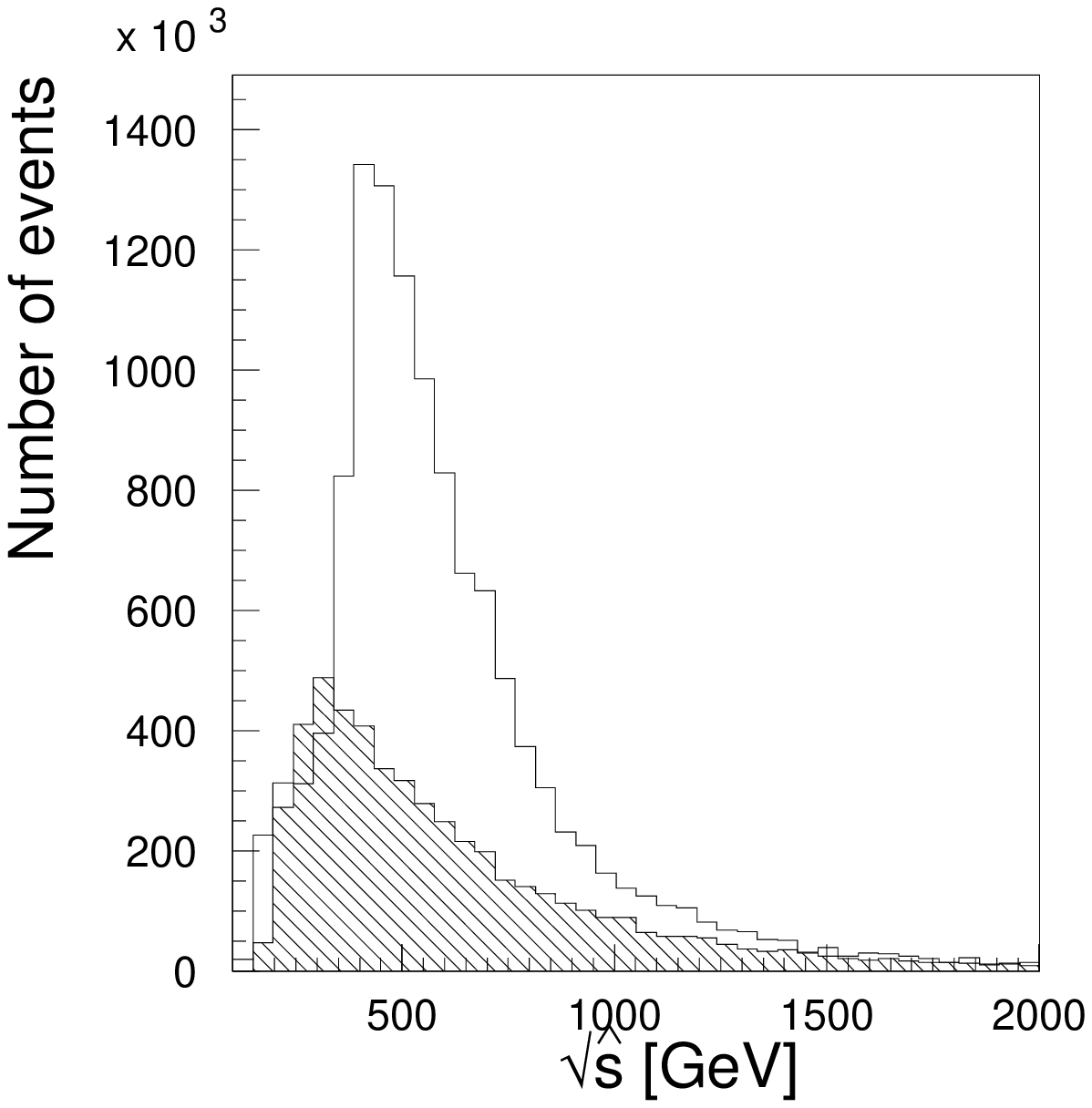}
    \vskip -1.8cm\hspace*{-0.5cm}
    \epsfxsize=8cm\epsffile{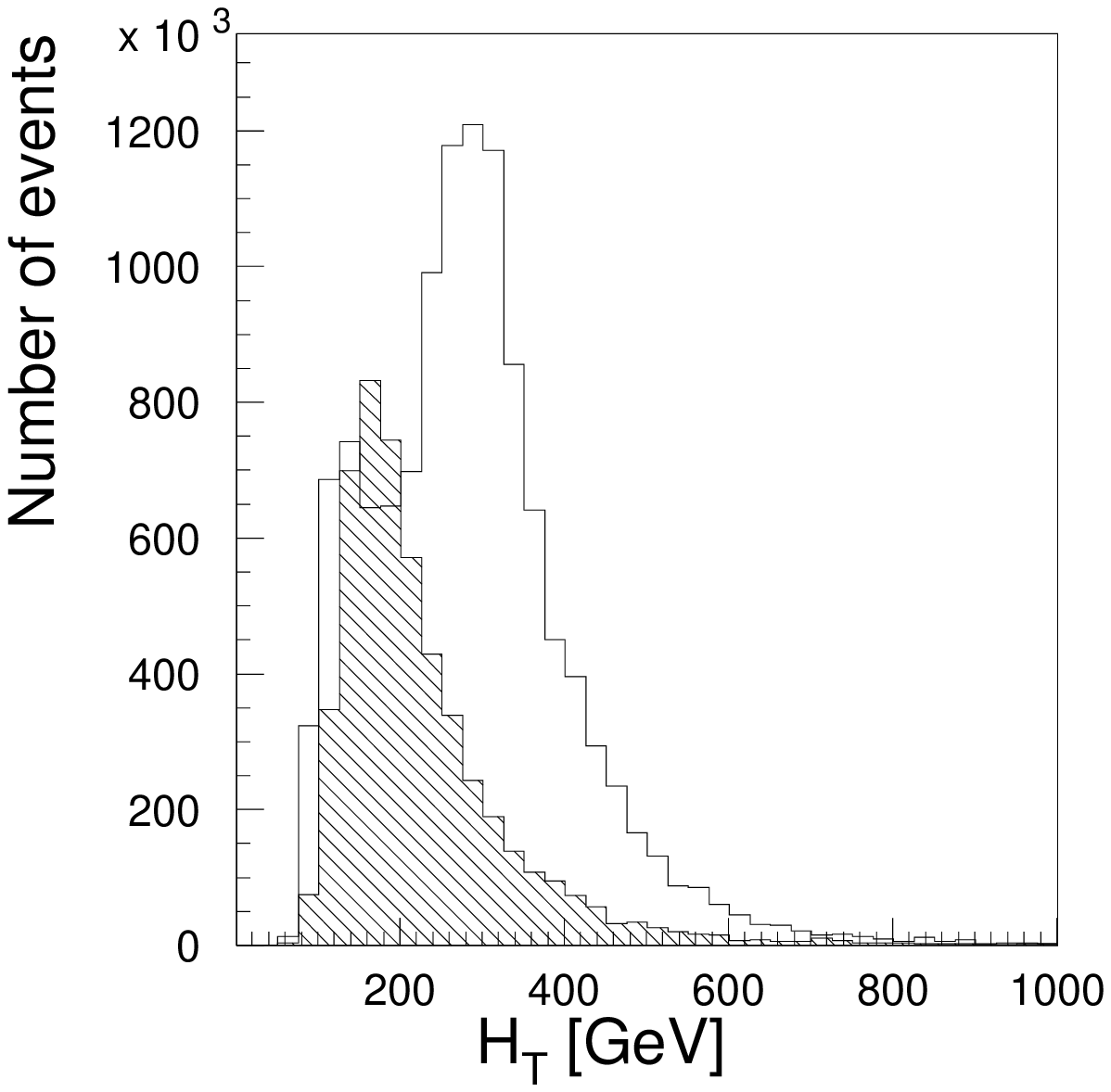}\epsfxsize=8cm\epsffile{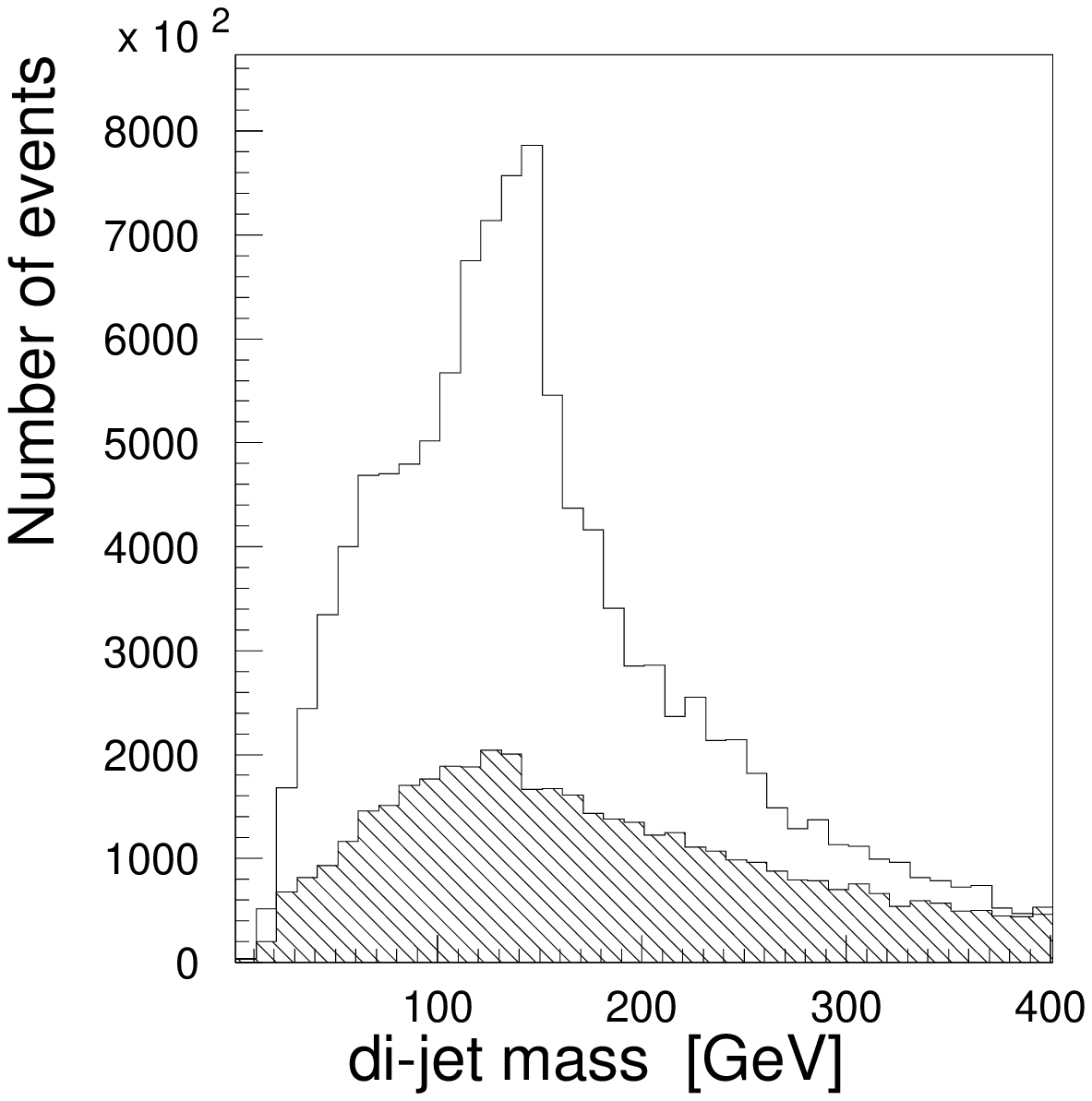}\\
  \end{center}
  \vspace*{-1.0cm}
  \caption{\label{st-fig_14} Distributions for signal and background
  for the some most spectacular  variables at LHC. Sketched histogram
  stands for signal.} 
  \vspace*{-0.5cm}
 \end{figure}

The distributions for  several sensitive  kinematical variables
for a separation of the signal and
the background are shown in Fig.~\ref{st-fig} for the Tevatron
and in Fig.~\ref{st-fig_14} for the LHC. The mentioned above 
effects of the jet fragmentation, detector resolution and  energy smearing
are included in the figures.
Among the kinematical variables for separation of the signal and
background the most attractive were found to be:
\begin{itemize}
\item $p_T$ of leading jet:\\
$p_T$ of leading jet distribution for the signal has a peak around $m_{top}/3$,
while it is much softer for QCD background at Tevatron(Fig.~\ref{st-fig}).
The main difference between  kinematical distributions for  signal
and background  at Tevatron is  that 
jets from $W+jj$ and $j(j)bb$  processes are 
 softer and less central than  those for signal with one very
hard jet coming from top and another softer  jet, accompanying top
quark. For LHC  there is no such striking
difference in $p_T$ of leading jet  distribution between signal and background.
It happens because of higher CM energy and dominating 
contribution to the background from $t\bar t$
production (Fig.~\ref{st-fig_14})
 \item $\sqrt{\hat{s}}$ - invariant mass of the 
system (Fig.~\ref{st-fig},\ref{st-fig_14}):\\
it is always bigger for signal than those for $Wjj$ and $j(j)bb$ backgrounds.
It is important to notice that $t\bar t$ background peaks at higher values of
invariant mass of the system which is  clearly seen in the case of LHC, 
where this background dominates.
\item $p_T W$: $W$ boson tends to be harder  from top-quark decay
than from QCD processes.
\item scalar transverse energy $H_T$ (Fig.~\ref{st-fig}c,\ref{st-fig_14}c ),
$H_T = |E_T({\rm{jet1}})| + |E_T({\rm{jet2}})| + |E_T({\rm{lepton}})|$ :\\
this kinematical variable has peak around 150 GeV for signal, around 300 GeV
for $t\bar t$ background and peaks at the small values for QCD background.
\item  di-jet mass (Fig.~\ref{st-fig},\ref{st-fig_14}):\\
 It is harder for signal
than for QCD background, for which $b\bar b$-pair coming mainly 
from gluon splitting,
in the same time di-jet mass distribution of $t\bar t$ background 
has similar shape with the signal.
Di-jet mass cut is also  used for a reduction of WZ background.
\end{itemize}


In our analysis we  used "effective" invariant top quark mass
variable which is constructed using the following algorithm.
It is  clear that mass of the top quark decaying to lepton, neutrino and b-quark
can not be unambiguously reconstructed since z-component 
of neutrino can not be measured.
One can construct top quark mass
\begin{eqnarray}
m_t^2 & = &\left(P_e + P_{\nu} + P_b \right)^2
\end{eqnarray}
using, one of two solutions for $p_{z\nu}$ of  simple quadratic equation
\footnote{Such a  method of the 
single  top quark mass 
reconstruction is known and has been used in the past
(see, C.-P.~Yuan, Phys. Rev.~D~{\bf 41}, 42 (1990))}
\begin{eqnarray}
m_W^2 & = &(P_e + P_{\nu})^2=80.12^2
\end{eqnarray}
Our Monte Carlo analysis shows that
if one  chooses $p_{z\nu}$ to be the  $|p_{z\nu}|_{min}$ from two solutions 
than it will be  in about $70\%$  true $p_{z\nu}$. 
In fact the reason for that is obvious and related to the fact that
smaller values of $p_{z\nu}$ correspond in most cases to
smaller values of the total invariant mass $\sqrt{s}$. And for 
the smaller values of $\sqrt{s}$ the effective parton-parton luminosity is 
larger and therefore the cross section is higher. 
However anyway if the one solution for $p_{z\nu}$ is used the invariant
"effective" mass distribution is broader than real invariant 
top quark mass and one should apply  rather wide 
window for this kinematical variable in order not to lose too much signal
events. We chose $\pm$50 GeV window in our analysis.

Based on  such  different behaviors of the signal and background
kinematical distributions 
the following set of cuts for the background suppression has
been worked out:
\begin{eqnarray}
&&\parbox{16cm}{
Cut 1: $\Delta R_{jj(ej)} >0.5$, $p_T jet > $ 10 GeV, $\mbox{$\rlap{\kern0.25em/}E_T$}>15$ GeV, ${p_t}_e>15$ GeV for Tevatron
\\
and $\Delta R_{jj(ej)} >0.5$, $p_t jet > $ 20 GeV, $\mbox{$\rlap{\kern0.25em/}E_T$}>20$
 GeV, ${p_t}_e>20$ GeV for LHC\\
which are "initial" cuts for jet separation and $W-$boson identification\\
Cut 2:  $p_t jet_{max }> 45$ GeV\\
Cut 3:  $\sqrt{\hat{s}} > $180 GeV \\
Cut 4:  $p_T W > $30 GeV\\
Cut 5:  di-jet mass$ >$ 25 GeV\\
Cut 6:  $H_T > $100 GeV for Tevatron and 260 GeV $>H_T>$ 100 GeV for LHC\\
Cut 7:  $3\ge$n-jet$\ge 2$\\
Cut 8:  di-jet mass $\ge 40$ GeV
}
\label{cuts}
\end{eqnarray}

The effect  of the consequent application of 
this  set of cuts is  presented in Table~\ref{st-tab},\ref{st-tab_14}.  
We should stress that "effective" mass window cut $\pm$50 GeV
was an initial cut and has been applied along with all others  cuts shown in the table.

Number of events 
presented in the tables as well as in the
Figs.~\ref{st-fig},\ref{st-fig_14} corresponds to the total
integrated luminosity 
2 fb$^{-1}$ (100~fb$^{-1}$)
\footnote{The numbers for the LHC could be easily rescaled to 
the 30 ~fb$^{-1}$ of the low luminosity LHC operation.}
for Tevatron (LHC) under the mentioned above assumptions of double
$b$-tagging efficiency 50\% and $b$-quark mistagging probability
0.5\%. 
{\small
\begin{table}[htb]
\begin{center}
\begin{tabular}{ | l | l |l | l | l | l | l  | l |  }
\hline
cuts  & signal & $Wb\bar{b}$& $Wjj$ & $WZ$ &  $j(j)b\bar{b}$ & $t\bar{t}$ & $WH$ \\
\hline
Cut 1   & 1.986$\cdot 10^2$ & 3.680$\cdot 10^2$ & 2.644$\cdot 10^2$ & 2.059$\cdot 10^1$  
	& 6.292$\cdot 10^2$ & 5.849$\cdot 10^2$ & 8.428$\cdot 10^0$\\
Cut 2   & 1.514$\cdot 10^2$ & 1.711$\cdot 10^2$ & 1.034$\cdot 10^2$ & 1.136$\cdot 10^1$  
	& 1.114$\cdot 10^2$ & 4.898$\cdot 10^2$ & 6.491$\cdot 10^0$\\
Cut 3   & 1.493$\cdot 10^2$ & 1.453$\cdot 10^2$ & 9.211$\cdot 10^1$ & 1.053$\cdot 10^1$  
	& 1.030$\cdot 10^2$ & 4.898$\cdot 10^2$ & 6.278$\cdot 10^0$\\
Cut 4   & 1.295$\cdot 10^2$ & 1.173$\cdot 10^2$ & 7.687$\cdot 10^1$ & 8.564$\cdot 10^0$  
	& 8.910$\cdot 10^1$ & 4.191$\cdot 10^2$ & 5.145$\cdot 10^0$\\
Cut 5   & 1.286$\cdot 10^2$ & 1.107$\cdot 10^2$ & 7.488$\cdot 10^1$ & 8.515$\cdot 10^0$  
	& 8.353$\cdot 10^1$ & 4.186$\cdot 10^2$ & 5.124$\cdot 10^0$\\
Cut 6   & 1.249$\cdot 10^2$ & 1.038$\cdot 10^2$ & 6.649$\cdot 10^1$ & 8.087$\cdot 10^0$  
	& 6.961$\cdot 10^1$ & 4.185$\cdot 10^2$ & 5.013$\cdot 10^0$\\
Cut 7   & 1.247$\cdot 10^2$ & 1.031$\cdot 10^2$ & 6.649$\cdot 10^1$ & 7.419$\cdot 10^0$  
	& 4.455$\cdot 10^1$ & 1.055$\cdot 10^2$ & 4.562$\cdot 10^0$\\
Cut 8   & 1.216$\cdot 10^2$ & 8.867$\cdot 10^1$ & 6.141$\cdot 10^1$ & 7.266$\cdot 10^0$
        & 3.619$\cdot 10^1$ & 1.039$\cdot 10^2$ & 4.490$\cdot 10^0$\\
\hline
\multicolumn{8}{|c|}{Signal: 122,  Background: 297; S/B$\simeq$ 0.41}\\
\hline
\end{tabular}
\end{center}
\caption{\label{st-tab}Number of events for single top signal and background at Tevatron.
Cuts numbering  correspond to (\ref{cuts}) set of cuts with their consequent application.
Window $\pm 50$ GeV around 175 GeV bin was imposed for reconstructed "effective" top mass.}
\end{table}
}
 
{\small
\begin{table}[htb]
\begin{center}
\begin{tabular}{ | l | l |l | l | l | l  | l | l |   }
\hline
cuts  & signal & $Wb\bar{b}$& $Wjj$ & $WZ$ &  $j(j)b\bar{b}$ & $t\bar{t}$ & $WH$ \\
\hline
Cut 1  & 1.212$\cdot 10^6$ & 8.236$\cdot 10^4$ & 1.724$\cdot 10^5$ & 1.912$\cdot 10^4$  
       & 1.155$\cdot 10^6$ & 4.449$\cdot 10^6$ & 6.124$\cdot 10^3$\\
Cut 2  & 8.792$\cdot 10^5$ & 5.143$\cdot 10^4$ & 1.058$\cdot 10^5$ & 1.177$\cdot 10^4$  
       & 6.112$\cdot 10^5$ & 3.762$\cdot 10^6$ & 4.923$\cdot 10^3$\\
Cut 3  & 8.764$\cdot 10^5$ & 4.871$\cdot 10^4$ & 1.015$\cdot 10^5$ & 1.138$\cdot 10^4$  
       & 6.053$\cdot 10^5$ & 3.762$\cdot 10^6$ & 4.854$\cdot 10^3$\\
Cut 4  & 7.423$\cdot 10^5$ & 3.826$\cdot 10^4$ & 7.758$\cdot 10^4$ & 9.048$\cdot 10^3$  
       & 4.974$\cdot 10^5$ & 3.262$\cdot 10^6$ & 3.976$\cdot 10^3$\\
Cut 5  & 7.401$\cdot 10^5$ & 3.771$\cdot 10^4$ & 7.735$\cdot 10^4$ & 9.013$\cdot 10^3$  
       & 4.957$\cdot 10^5$ & 3.262$\cdot 10^6$ & 3.972$\cdot 10^3$\\
Cut 6  & 5.643$\cdot 10^5$ & 3.649$\cdot 10^4$ & 7.524$\cdot 10^4$ & 7.545$\cdot 10^3$  
       & 4.729$\cdot 10^5$ & 6.214$\cdot 10^5$ & 3.334$\cdot 10^3$\\
Cut 7  & 5.370$\cdot 10^5$ & 3.610$\cdot 10^4$ & 7.408$\cdot 10^4$ & 6.122$\cdot 10^3$  
       & 2.411$\cdot 10^5$ & 1.886$\cdot 10^5$ & 2.740$\cdot 10^3$\\
Cut 8  & 5.296$\cdot 10^5$ & 3.177$\cdot 10^4$ & 7.019$\cdot 10^4$ & 6.030$\cdot 10^3$  
       & 2.301$\cdot 10^5$ & 1.886$\cdot 10^5$ & 2.694$\cdot 10^3$\\
\hline
\multicolumn{8}{|c|}{Signal: $5.3\cdot 10^5$,  Background: $5.3\cdot 10^5$ ; S/B$=$ 1.0}\\
\hline
\end{tabular}
\end{center}
\caption{\label{st-tab_14}Number of events for single top signal and background at LHC.
Cuts numbering  correspond to (\ref{cuts}) set of cuts with their consequent application.
Window $\pm 50$ GeV around 175 GeV bin was imposed for reconstructed "effective" top mass.}
\end{table}}

>From the tables one can see that in fact two  cuts,
Cut 2 reducing  the $QCD+Wjj$ background
and Cut 7 eliminating $t\bar{t}$ background, play 
the leading role. In the same time all cuts are strongly 
correlated  and one can effectively replace Cut 2 by Cut3+4 or 
more complicated combination with the same success.

\begin{figure}[htb]
   \begin{center}
   \begin{picture}(600,230)(0,0)
   \vspace*{-0.6cm}
   \hspace*{-0.5cm}
   \epsfxsize=8cm\epsffile{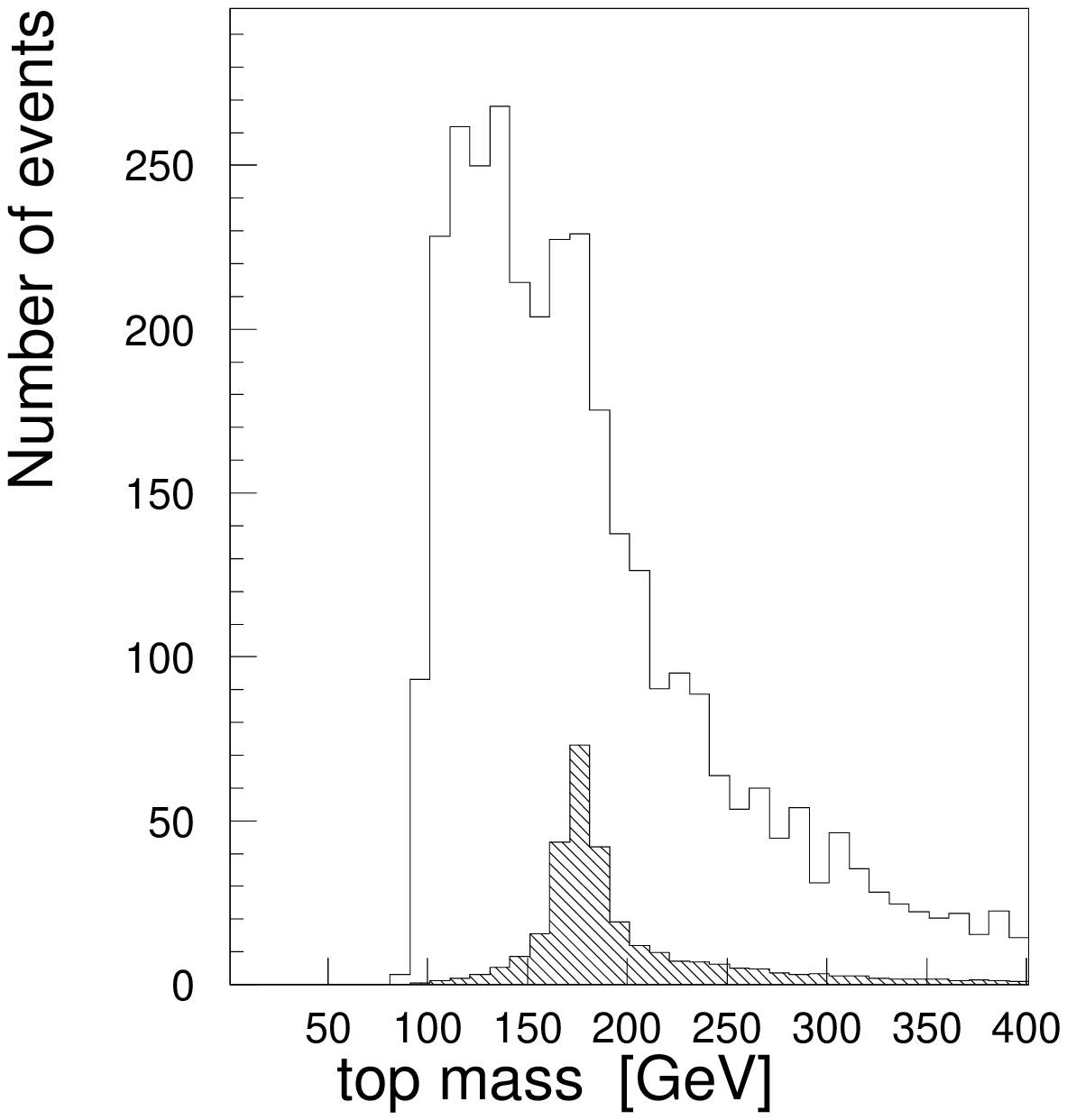}\epsfxsize=8cm\epsffile{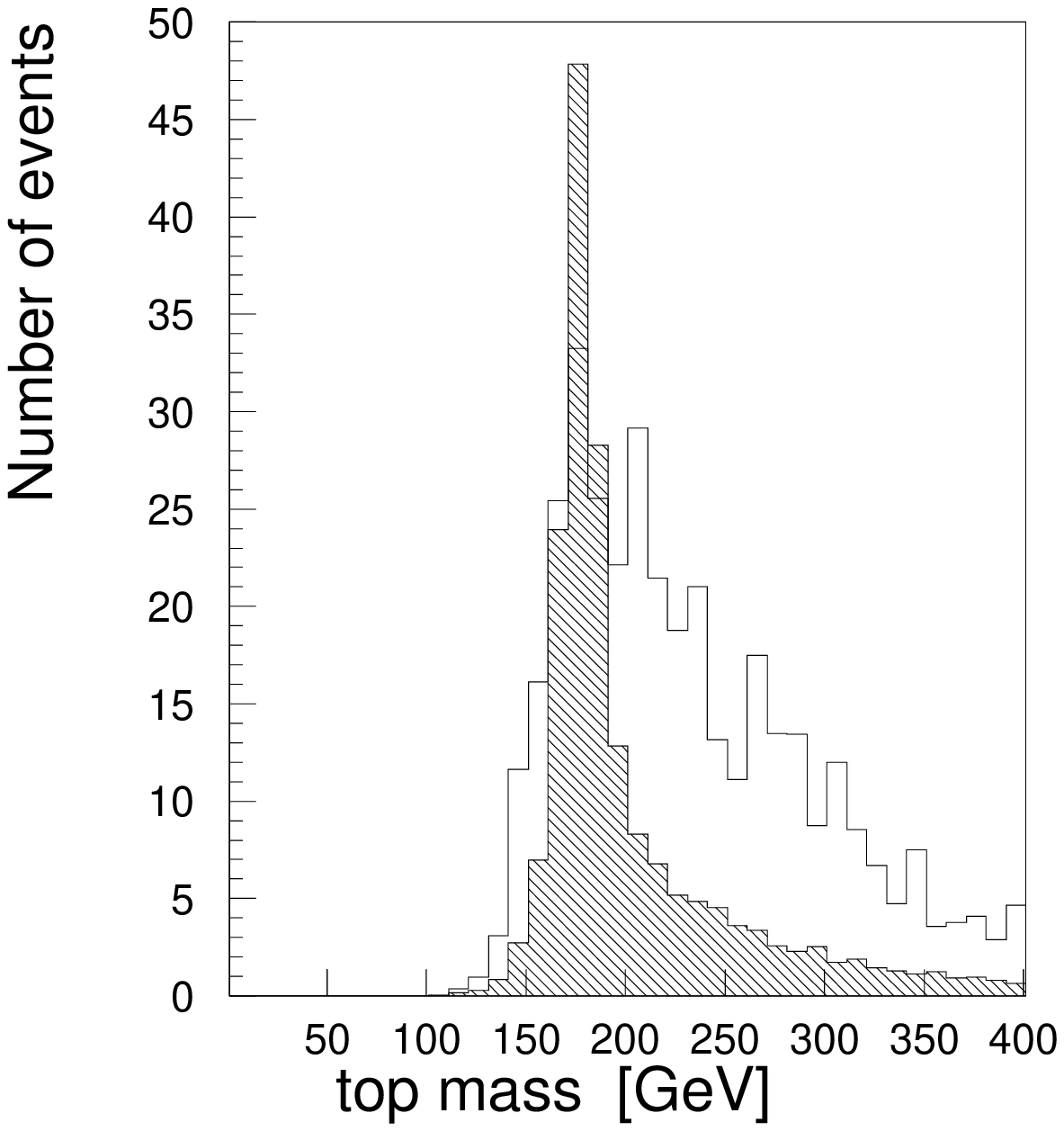}
         \put(-390,  230){(a)}
         \put( -150, 230) {(b)} 
   \end{picture}
   \end{center}
   \vspace*{-1.0cm}
   \caption{\label{st-fig2} Distributions for invariant top mass before (a) and after (b)
    cut application at Tevatron. Sketched histogram
  stands for signal.} 
   \end{figure}
\begin{figure}[htb]
   \begin{center}
   \begin{picture}(600,230)(0,0)
   \vspace*{-0.6cm}
   \hspace*{-0.5cm}
   \epsfxsize=8cm\epsffile{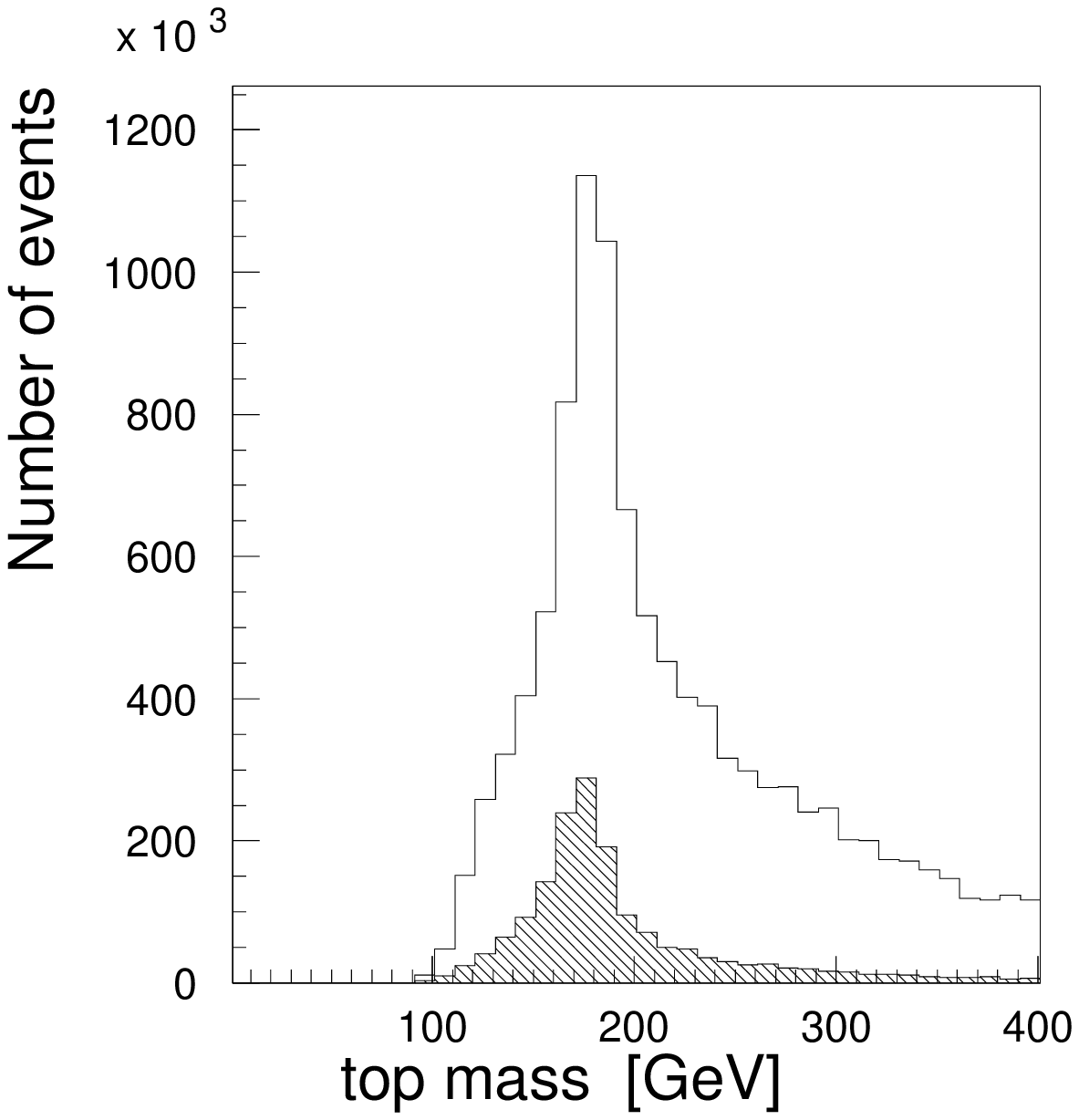}\epsfxsize=8cm\epsffile{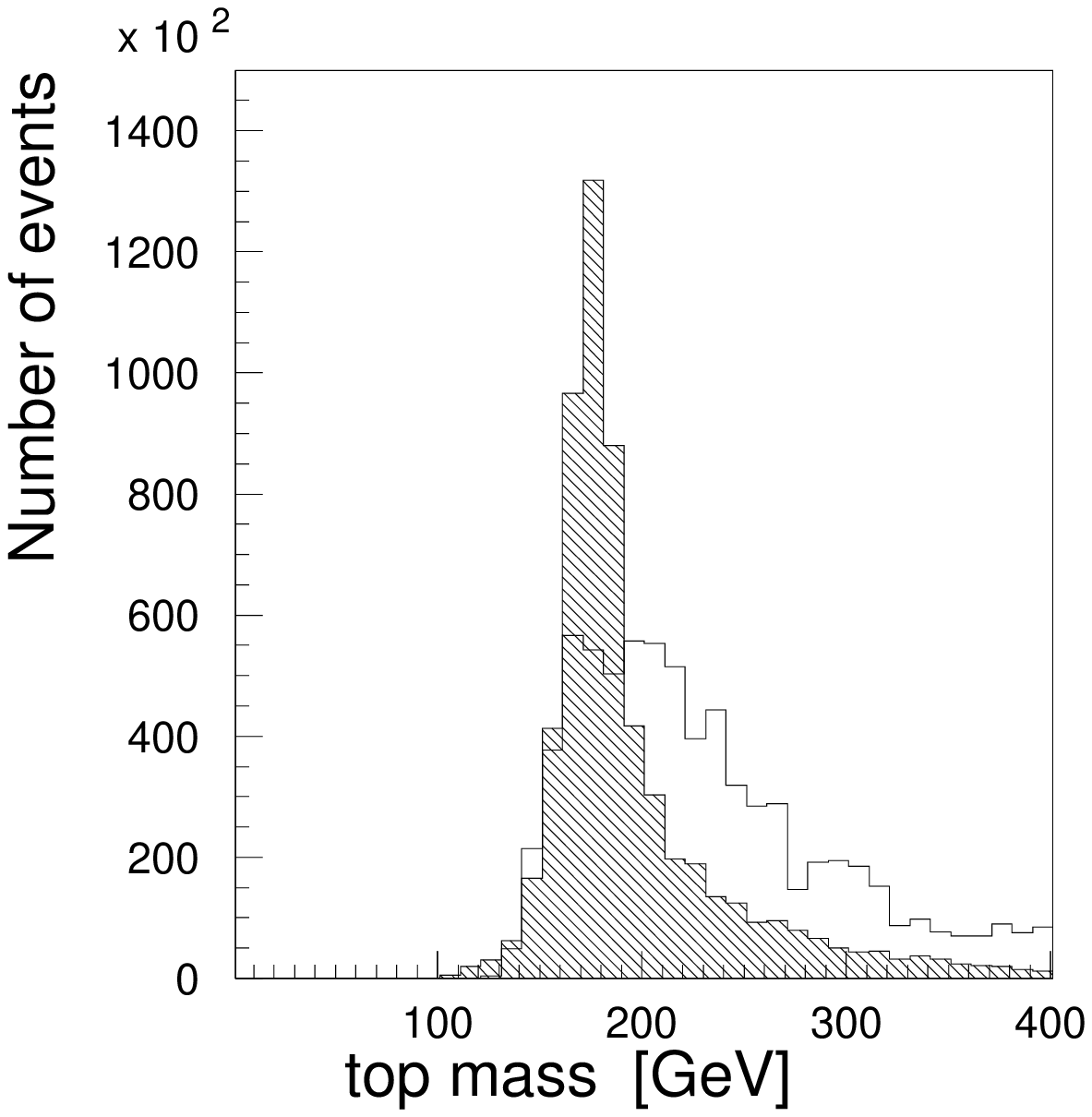}
         \put(-390,  210){(a)}
         \put( -150,  210){(b)} 
   \end{picture}
   \end{center}
   \vspace*{-1.0cm}
   \caption{\label{st-fig2_14} Distributions for invariant top mass before (a) and after (b)
    cut application at LHC. Sketched histogram
  stands for signal.} 
\end{figure}

The strong background reduction is clearly illustrated in
Fig.~\ref{st-fig2}a,b,\ref{st-fig2_14}a,b  
for the  invariant top mass distribution
before (a) and after (b) application of kinematical cuts.
After cuts applied the background  
became about 10 times smaller at the Tevatron and 18 times
at the LHC while 
approximately 60\% (40\%) of signal survived at Tevatron (LHC). 
Signal/background
ratio becomes equal approximately to  0.4 at Tevatron and  1 at LHC.
Such a background suppression will allow
to measure the signal cross section with the high accuracy.


The cross section for single top quarks includes the $Wtb$
coupling directly, in contrast to {\mbox{${t\bar{t}}$}} pair production.
Therefore, single top production provides a unique opportunity to
study the $Wtb$ structure and to measure {\mbox{$V_{tb}$}}. Experimental
studies of this type are among the main goals of the single top
physics. 
Using the single top quark search one 
 can examine the effects  of a deviation in the $Wtb$
coupling from the SM structure and directly measure 
 the CKM matrix element {\mbox{$V_{tb}$}}.
Since the signal to background ratio is high after kinematical cuts applied
the error of {\mbox{$V_{tb}$}} measurement as was shown in \cite{our-st} 
is expected to be of
order of 10\% at the Tevatron RUN2.
In the same time  much higher statistics and  
good signal/background ratio at LHC
allow considerably improve the measurement of {\mbox{$V_{tb}$}}  value 
and test $Wtb$ vertex.
Since statistical error for  $10^5$ events is less then 1\%, then uncertainty 
of $Wtb$ vertex measurement at LHC depends mostly on  the
uncertainty of theoretical calculations for  
single top quark production cross section and for the backgrounds.
That is why calculations of the next order   
corrections to the single top quark production 
including the corrections to the kinematical distributions
but not only to the total events rate and a simulation of main backgrounds
at the NLO level are important for the LHC.  

 Another important source of uncertainties in the $Wtb$ vertex
measurement comes from parton distribution uncertainty as well as
from the accuracy of top-quark measurement.
In case of Tevatron these uncertainties have been included
into consideration \cite{our-st}. However in case of LHC
 this point is not very clear since one does not know how large 
 those uncertainties would remain  when the  experiment 
 will start, and the parton distribution functions and the top mass
will be measured in separate experiments. 
That is why at present stage we did not include the pointed uncertainties
for the case of LHC.  

\section{Conclusions}
The study of the single top-quark production  versus complete 
background processes
has been done. For calculations a special generator has been
created  based on the CompHEP and PYTHIA/JETSET programs. 
The computation  shows
the importance of the QCD fake background which was not taken into account
in the previous papers. Study of the effects of the initial and 
final state radiation for $jbb$ process shows that such an
approximative method of simulation of higher jet multiplicity process
has the accuracy of order 10\% or less for the rate and gives
significantly softer $p_T$ distribution of the radiating jet 
comparing to the complete calculations. 

It was shown that after optimized cuts applied the signal from the single
top quark can be extracted from the background with the
signal to background ratio about
0.4 for the upgraded Tevatron and  1 for the LHC. 
The remaining after cuts single top rate in
the $lepton+jets$ mode is expected 
to be about 120 events at the upgraded Tevatron
and about $1.6\cdot10^5$ events at the low luminosity LHC operation with
30 $fb^{-1}$ accumulated data and assumptions
made  above. 
One can expect that Vtb CKM matrix element can be measured at 
upgraded Tevatron with an accuracy about 10\% and hopefully 
with an accuracy of the order of few \% at LHC.

\subsection*{Acknowledgments}
Authors are grateful to members of the single top group 
of the D0 collaboration for useful discussions.

A.\ B.\ is grateful to S. F. Novaes for fruitful discussions, thank 
the Instituto de
F\'{\i}sica Te\'orica for its kind hospitality and acknowledges support
from Funda\c{c}\~ao de Amparo \`a Pesquisa do Estado de S\~ao Paulo
(FAPESP). 

E.\ B.\ would like to thank H.~Anlauf, P.~Manakos, T.~Ohl, A.~Pukhov,
V.~Savrin, J.~Smith, C.-P.~Yuan, and B.-L.~Young for discussions of
different aspects of calculations used, 
he wishes to acknowledge the
KEK Minami-Tateya (GRACE) collaboration for the kind 
hospitality during his visit at KEK and his colleagues
from the CompHEP group for the interest and support.

E.\ B.\ and L.\ D.\ acknowledge 
the financial support of the Russian Foundation of Basic Research  
(grant No 96-02-19773a), the Russian Ministry of Science and
Technologies, and the Sankt-Petersburg Grant Center.

\end{document}